\documentclass[preprint]{aastex}
\usepackage{graphicx}
\usepackage{url}
\begin{document}
\title
{The Synergy of Direct Imaging and Astrometry \\for Orbit Determination of exo-Earths}
\author
{Michael Shao, Joseph Catanzarite and Xiaopei Pan}

\affil{ Jet Propulsion Laboratory, California Institute of Technology \break
4800 Oak Grove Dr, Pasadena, CA 91109, USA  }
\email{ michael.shao@jpl.nasa.gov, joseph.catanzarite@jpl.nasa.gov \\ and xiaopei.pan@jpl.nasa.gov}

\begin{abstract}
The holy grail of exoplanet searches is an exo-Earth, an Earth mass planet in the habitable zone around a nearby star. Mass is one of the most important characteristics of a planet and can only be measured by observing the motion of the star around the planet-star center of gravity. The planet's orbit can be measured either by imaging the planet at multiple epochs or by measuring the position of the star at multiple epochs by space based astrometry.  The measurement of an exo-planet's orbit by direct imaging is complicated by a number of   factors. One is the inner working angle (IWA). A space coronagraph or interferometer imaging an exo-Earth can separate the light from the planet from the light from the star only when the star-planet separation is larger than the IWA. Secondly, the apparent brightness of a planet depends on the orbital phase. A single image of a planet cannot tell us whether the planet is in the habitable zone or distinguish whether it is an exo-Earth or a Neptune-mass planet. Third is the confusion that may arise from the presence of multiple planets. With two images of a multiple planet system, it is not possible to assign a dot to a planet based only on the photometry and color of the planet. Finally, the planet-star contrast must exceed a certain minimum value in order for the planet to be detected. The planet may be unobservable even when it is outside the IWA, such as when the bright side of the planet is facing away from us in a `crescent' phase.
In this paper we address the question: ``Can a prior astrometric mission that can identify which stars have Earthlike planets significantly improve the science yield of a mission to image exo-Earths?''   In the case of the Occulting Ozone Observatory (O3),  a small external occulter mission that cannot measure spectra, we find that the occulter mission could confirm the orbits of $\sim4$ to $\sim5$ times as many exo-Earths if an astrometric mission preceded it to identify which stars had such planets. In the case of an internal coronagraph we find that a survey of the nearest $\sim60$ stars could be done with a telescope 1/2 the size if an astrometric mission had first identified the presence of Earth-like planets in the habitable zone and measured their orbital parameters.

\end{abstract}

\section{Introduction }

\begin{figure}
\scalebox{0.8}{\includegraphics{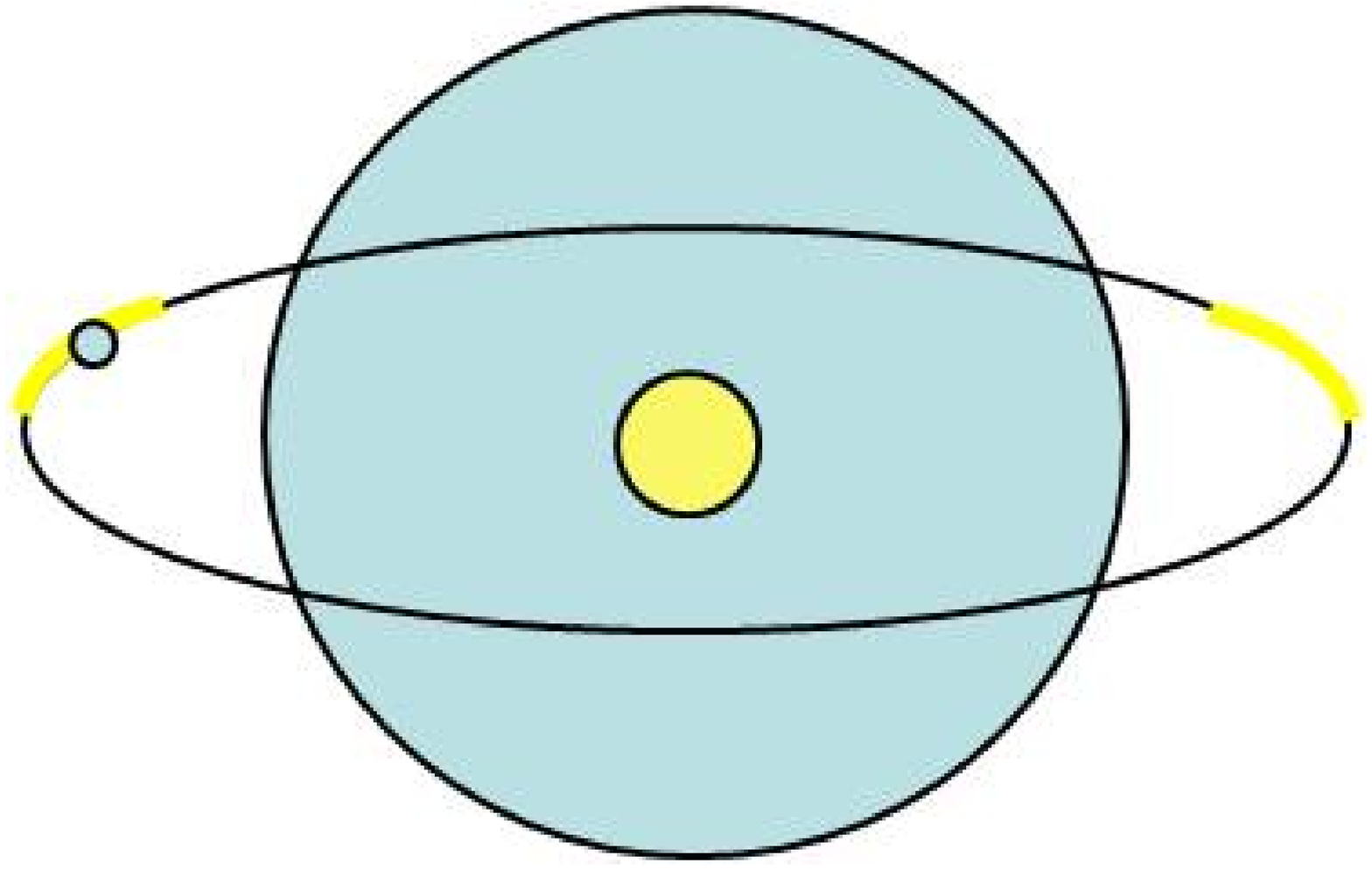}}

\caption{Observable IWA and planet orbit. A planet is observable only when it is outside the IWA and its flux contrast exceeds the minimum contrast.\label{fig:Figure1}}

\end{figure}

When a potential exo-Earth is detected, the first thing we want to know is, ``Is this an Earth?'' The question has three parts:  first ``Is the mass of the planet roughly the same as our Earth?''; second, ``Is the orbit of the planet in the habitable zone of the star?''; and third, ``What does the spectral characterization of the planet tell us?'' Direct imaging measures only the motion of the planet. Because the albedo, the phase, and surface scattering properties of the planet are unknown, planet mass cannot be determined from images. And while spectroscopic information may indicate whether the planet's atmosphere is qualitatively similar to that expected for a Neptune or a terrestrial planet, the planet's mass can only be quantitatively determined by astrometry -- measuring the motion of the star around the planet-star center of gravity.
Measuring the orbit of a planet can be done by either astrometric or direct imaging missions. But because of the limitation of the IWA an exo-Earth is generally observable over a smaller fraction of its orbit, for a direct imaging mission. Figure~\ref{fig:Figure1} shows the orbit of a planet (blue dot) around another star (represented by the yellow dot).  The large blue circular region is the IWA and the yellow arcs are the parts of the orbit when the planet is observable. With a coronagraph whose IWA is only slightly smaller than the maximum star-planet separation, orbital parameters such as inclination can't be measured. If a planet is non-self-luminous and its surface is a Lambertian scatterer, then for a circular orbit, the planet's apparent brightness can vary by a factor of three from the ``full moon'' phase to the ``half moon'' phase (see Equation~\ref{eq:Lambertian}). In multiple planet systems, two images with one planet in each image leaves open the possibility that they are two separate planets.
If the IWA is substantially smaller, say half of the maximum star-planet separation, the planet becomes observable over most of its orbit. In this case it will be possible to look for seasonal variations in brightness. Such variations may occur because the surface is a non-Lambertian scatterer. Another source of seasonal variation may be a change in albedo. In the winter time, an ocean's surface may be covered with ice, which has a much higher albedo than liquid water.   Seasonal changes in brightness are a double edged sword; they    reveal important information about the surface of the planet, but their presence also complicates the identification of which dot is which planet and the determination of the planet's orbit around the star.

The key to measuring the orbit of a planet is to have many images of the planet at different times of its year. But some types of coronagraphs are seriously limited in their ability to take images at many epochs. Recent studies have shown that an astrometric mission will have a completeness for detection of $\geq95\%$ for star-planet systems within 10 pc of the Sun \cite{tra10}.  On the other hand, the completeness for detection for a direct imaging mission, in the absence of data from an astrometric mission, has been shown to be $\leq35\%$ \cite{sav10}.  Given proof-of-existence, masses, and orbits from an astrometric mission, a direct imaging mission is free to plan and optimize its resources for the function for which it is uniquely-equipped, the spectral characterization of planetary atmospheres.  With this information in hand, the exoplanet's resum\'{e} --- existence, mass, orbit, and atmospheric spectra --- is complete.  Only then can we answer the question, ``Is this an Earth?''

In this paper, we examine and quantify the effect of confusion on imaging surveys.  We investigate the synergy between astrometry and direct imaging in measuring the orbit of a planet, and we examine how an astrometric orbit helps a direct imaging mission determine that a `pale blue dot' is an exo-Earth and not a false alarm. We consider only the case of circular orbits. We estimate the number of observations needed for orbit determination with imaging alone and with a combination of imaging and astrometry.  Using simulations based on reasonable assumptions, we show that knowledge from a prior astrometric survey can greatly increase the science yield of an imaging mission.
In sections 2 and 3 we consider two different situations. In section 2 the planet is first discovered by an astrometric mission and the role of imaging is to confirm the discovery and improve on its orbit determination. In section 3 we consider the measurement of an orbit from direct imaging alone. In section 4 we discuss how to realistically model planet populations.  We simulate an example external occulter mission, O3, and show that the number of exo-Earths with confirmed orbits is significantly different with and without a prior astrometric mission.  In section 5 we extend the discussion to internal coronagraphs and compute the time required to get orbits for a medium to large internal coronagraph mission.

\section{Planetary orbit determination from astrometry followed by imaging}

When we image a planet as a dot in a sea of speckles, we want to know if it is potentially habitable. We want to know its mass and the semi-major axis of its orbit. For an Earth twin in a 1 AU orbit at 10 pc, only astrometry at the sub-microarcsecond level can measure the mass of the planet with reasonable precision ($\pm 0.3M_{\oplus}$).  But both astrometry and imaging can in theory measure the orbit. Astrometry does this indirectly by measuring the reflex motion of the star.  There is no IWA limitation, and the motion of the planet as inferred from the reflex motion of the star can be measured throughout the orbit.

\subsection{Astrometric orbit precision}
Space-based narrow-angle astrometry (such as the SIM Lite mission) can detect 1$M_{\oplus}$ planets in the habitable zone of about 60 nearby stars in a five-year mission \cite{unw08}.  NASA conducted a double blind study for the astrometric detection of Earth-like planets in multiple planet systems \cite{tra10}. The result of the test was that the presence of multiple planets has a marginal to negligible impact on the astrometric mission's ability to detect and measure the orbits of terrestrial planets (planets of mass 1 to 10 $M_{\oplus}$) in the habitable zone. One of the side products of that study was a determination of the accuracy of the astrometric orbit at the ``edge'' of detectability. For a star at a distance d in pc, with a planet causing reflex motion in the star with orbital semimajor axis $a_{*}$ in AU (astronomical units), the astrometric signature in arcseconds is $\alpha = a_{*}/d$.

If N differential measurements of the star's position are taken along each of two orthogonal axes, with single-measurement accuracy $\sigma$ , then the Signal to Noise Ratio (SNR) can be defined as
\begin{equation}
SNR=\frac{\alpha}{\sigma} \sqrt{\frac{N}{2}}
\end{equation}

From extensive Monte Carlo studies of periodograms, we determined that a mission with a SNR of at least 5.8 was necessary to detect planets with a false alarm probability of only 1\%.
Assuming a circular orbit with period P $\ll$ mission duration T, and even sampling, the Cramer-Rao bounds on the 1-$\sigma$ errors in mass and period give fractional errors of
\begin{equation}
	\frac{\sigma_{m}}{m}=\frac{\sqrt{2}}{SNR}
\end{equation}
and \begin{equation} \frac{\sigma_{P}}{P}=\frac{\sqrt{6}P}{\pi T} \frac{1}{SNR}.
\end{equation}

At SNR of 5.8, the period of a 1 year planet would have a 1-$\sigma$ error of 3\% and for a $1M_{\oplus}$ planet, the 1-$\sigma$ mass error would be $0.3M_{\oplus}$. In indirect detection, the semi-major axis of the orbit is derived from its period using Kepler's laws.

\begin{figure}
\scalebox{0.8}{
\includegraphics{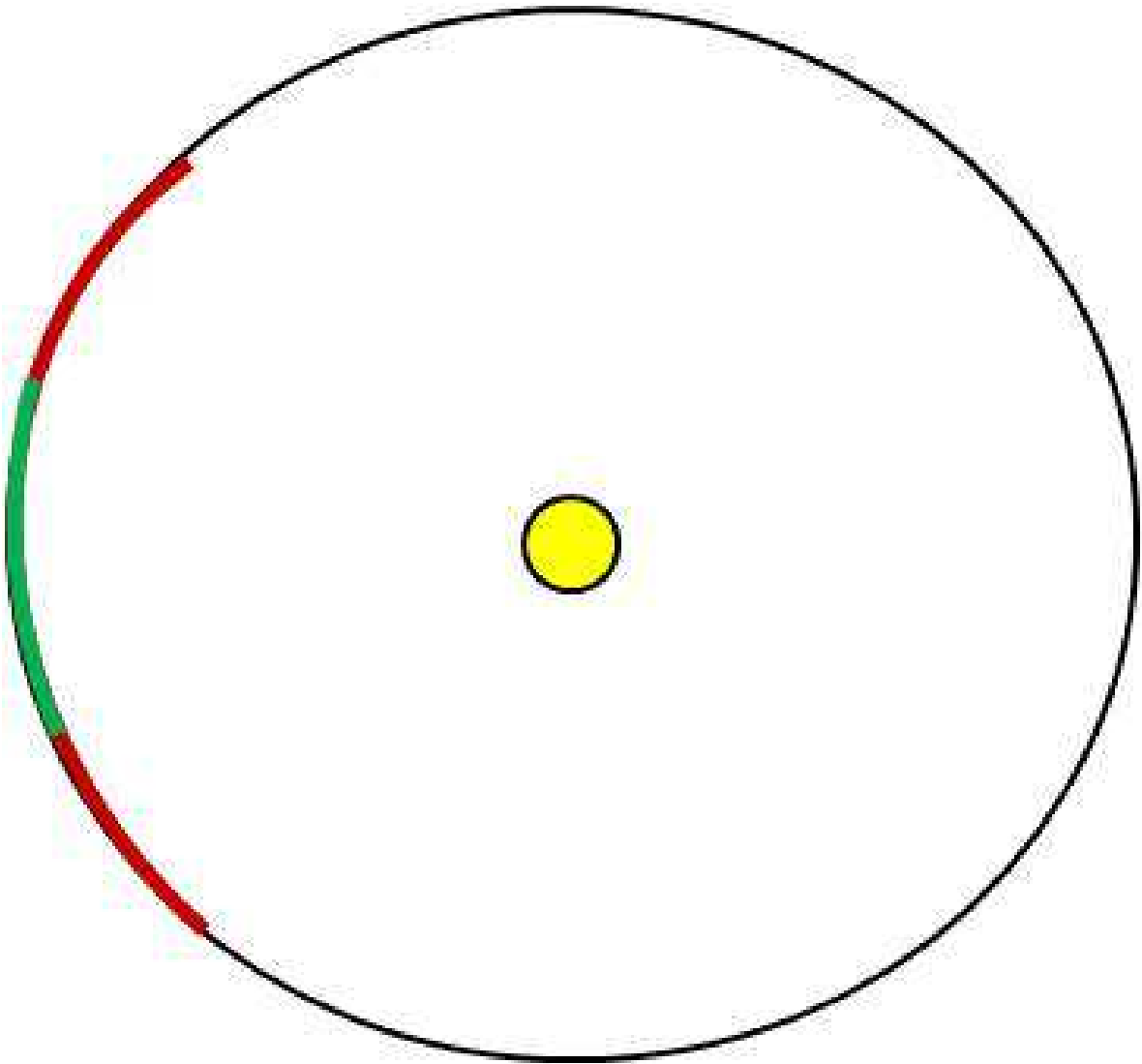}
}
\caption{Astrometric Orbit Error. The error in phase grows with time, from $\pm14$ days at mid-epoch (green) to $\pm50$ days 5 years after mid-epoch (red).\label{fig:Figure2}}
\end{figure}

Under the same assumptions, the Cramer-Rao bound on the 1-$\sigma$ error in orbital phase is given by
\begin{equation}  \sigma_{\Phi}=\frac{\sqrt{\frac{24(t-t_{0})^2}{T^2}+2}}{SNR},
\end{equation}
where t is the time and $t_{0}$ is the time at the midpoint of the astrometric mission. For a 5 year mission, the orbital phase at mid-mission has an error of roughly 0.24 radians ($\pm14$ days in a 365 day year).  If the astrometric data preceded the imaging search by 5 years, the uncertainty in the orbital period would cause the orbital phase error bar to grow linearly with time.  Five years after the mid-epoch of the astrometry data, the orbital phase uncertainty would be roughly $\pm50$ days. An exo-Earth at 10 pc is visible for $\pm38$ days on each of the two lobes of its orbit (see Figure~\ref{fig:Figure2}), for an instrument with IWA of 75 mas and minimum contrast of 4$\times10^{-11}$ and an orbit inclination of $80\degr$.

We consider only circular orbits in this paper; the uncertainty in the orbital phase would depart from linearity as the orbit becomes more eccentric. All other things being equal, a moderately  eccentric orbit (as compared to a circular orbit) has more uncertainty, whether the position of the star is measured by astrometry, or the position of the planet is measured from imaging.  This is because an eccentric orbit needs two  additional parameters: eccentricity, and  periastron angle. There is some degeneracy between inclination and eccentricity, so that a highly inclined orbit may mimic an eccentric one. It is important to measure the eccentricity of the orbit of a potential exo-Earth to gauge its habitability. If the orbit is eccentric the planet can travel inside and outside the habitable zone, even though its orbital semimajor axis is within the habitable zone.

\subsection{Imaging verification and refinement of the orbit}

If we start an imaging search of an Earth-Sun twin at 10 pc that was previously found astrometrically, the astrometric error bar 5 years after the mean epoch of astrometric measurements would be 0.03 AU in the radial direction and about 0.85 AU in the circumferential direction.   A single image of the planet taken five or more years after the astrometric data set would substantially reduce the error bar in the circumferential direction. With this knowledge, we can accurately predict when the planet will be observable in the future. An additional image would then confirm the planet's existence and further refine the orbit solution.

If the coronagraph were working at $2\lambda/D$, and planet-star separation of 1 AU corresponded to  $2\lambda/D$, then with a 2 meter telescope, a single SNR=5 image in the optical band would locate the planet to  roughly 0.1 AU. The major error in the astrometric orbit is in the circumferential position of the planet ($\sim1$ AU), due primarily to the 5 year time delay between the astrometric survey and imaging followup. At the time the first image is taken, the orbital phase uncertainty would be 0.1 AU, degrading to 0.15 AU 2 years later. So the circumferential error is significantly reduced by one image. The one image would also reduce the period uncertainty from 3\% to 0.9\%.  With 5 prior years of astrometric data, there should be no problem finding the planet a second time.  This is in contrast to orbit determination with imaging data alone, as explained in the next section.

\section{Planetary orbit determination from imaging alone}

When a planet orbits a star, a coronagraph detects light from the star that is reflected by the planet. The flux contrast of the planet is determined by the distance from the star and the angle $\alpha$ from the star to the planet to the observer. For a circular orbit, if {\it i} is the inclination and $ \phi $ is the orbital phase, then $cos(\alpha)= -sin(\phi)  sin(\it i)$. Assuming the planet is a Lambertian scatterer the planet-star flux contrast is given by
\begin{equation}\label{eq:Lambertian}
    F(\phi)=ap\left(\frac{r}{R(\phi)}\right)^{2} \frac{sin(\alpha)+(\pi-\alpha)cos(\alpha)}{\pi},
\end{equation}
where $R(\phi)$ is the distance from the star to the planet, {\it r} is the radius of the planet, {\it a} is the planet's albedo, and p is 2/3 for a Lambertian scatterer. As the planet orbits the star it undergoes `phases' as $\alpha$ changes periodically. An important consideration is that a planet with uniform albedo in reflected light will exhibit a factor of 3 change in apparent brightness from zero phase ``full moon'' to $90\degr$ phase ``half moon''. Therefore, if the system has multiple planets, the apparent brightness of an image cannot be used to identify a planet.

In principle, three images of one planet at 3 different epochs are sufficient to determine an orbit. But since we can't use photometry to assign a dot in an image to a specific planet, a fourth detection of the planet is needed to ensure that all 4 observations were of the same planet \cite{gla07}. If the planet is observable only over 20\% of the orbit, getting an orbit requires, on average, 20 images.
Some types of coronagraphs, such as the external occulters, allow a very limited number of visits. External occulters use a large ($\sim50$ m) star shade at a long distance (tens of thousands of km)  in front of a telescope to block the starlight. The preliminary designs of  occulter missions allow 120 to 130  observations over 5 years \cite{gla07, lin07}. The limited number of visits is not sufficient, without a prior astrometric mission, to obtain orbits of  more than a few candidate exo-Earths \cite{sav10}.  Even if a star has just one planet, one or two images of the system cannot establish whether or not the star has a  planet in the habitable zone. Consequently, one has to continue taking images of the system until there is conclusive evidence that a planet is either present or absent in the habitable zone.  Four detections are needed to measure and confirm a planet's orbit, and unless the entire orbit is outside the IWA, it takes more than four tries to get four detections.  An imaging survey cannot measure the mass of a planet, and thus cannot determine whether or not a planet is a terrestrial planet.  A planet determined to be in the habitable zone by an imaging mission will still have an unknown mass.
Section 4 examines a small occulter mission with and without a prior astrometric mission in a detailed comparison.

\subsection{Observational requirements and number of target stars}

\begin{figure}
\includegraphics{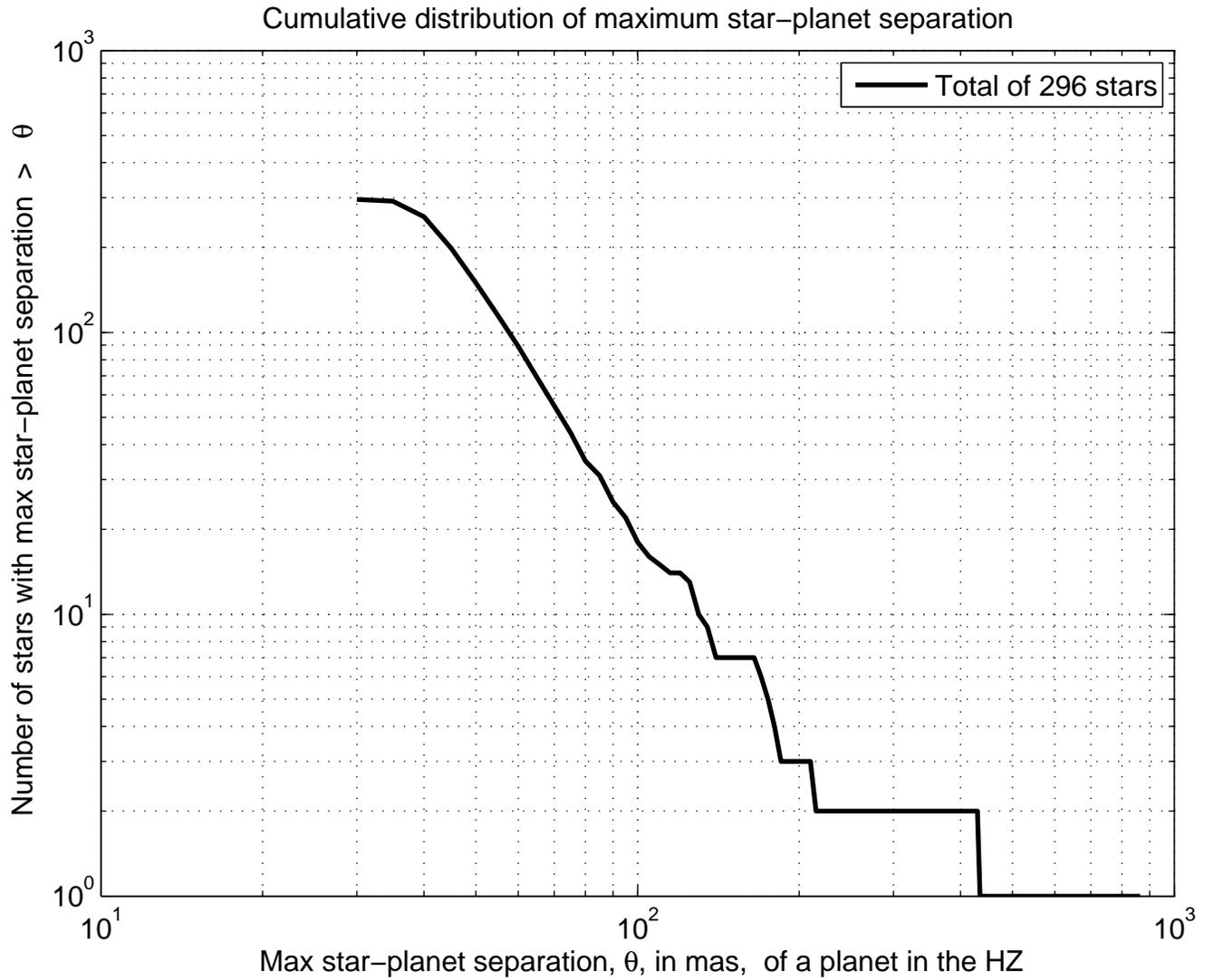}
\caption{Cumulative distribution of the number of stars for which an exo-Earth at mid-habitable zone is separated from its star by more than $\theta$.  There are 296 stars with planet separation $> 30$ mas.\label{fig:Figure3}}

\end{figure}

The Exo-Planet Task Force (ExoPTF) report recommended that an astrometric mission be able to survey 60 to 100 nearby stars for Earth-like planets \cite{lun08}. We feel that this is an appropriate number for a direct imaging mission as well.
Until Kepler data are analyzed and followed up, the fraction of stars with terrestrial planets in the habitable zone remains unknown, and we can only make an educated guess. If we look at the fraction of stars that have Jovian planets, we find that the number of planets (per unit mass) increases dramatically at low masses.  A plot of density vs. log(Mass) and log(Period) shows that the density  varies slowly with log(M) and log(P).  From periods of 3 days to 3 years and masses from 0.3 to 10 $M_{Jupiter}$, about 8\% of stars have planets and about 10\% of stars that have planets have multiple Jovian planets \cite{cum08}. If the densities in log (M) and log (P) are extrapolated, about 1\% of stars would have terrestrial planets in the habitable zone. The reason for the small number is the small  phase space volume of the habitable zone. More recently, the Swiss RV (Radial Velocity) team has stated that for orbit periods between 3 days and 3 months and masses between 5 to 50 Earths, up to 30\% of stars have one such Neptune/super-Earth type planet \cite{may09a}. Compared to the known density of the gas-giants, this is a dramatic increase in  density. Recent models of core accretion theory predict the rise in the density of low-mass planets and show remarkable consistency with the RV observations \cite{mor09, ida08}. When extrapolated to terrestrial planets of 1 to 10 $M_{\oplus}$ in the habitable zone, $\sim10\%$ of stars are expected to have such planets.
While we won't have data on the prevalence of exo-Earths in the habitable zone until Kepler data have been analyzed, an estimate of 10\% for $\eta_{\oplus}$ seems a reasonable guess given current knowledge. A coronagraph capable of detecting an exo-Earth in the habitable zone of 60 nearby stars seems like a reasonable ``minimum'' capability for a mission designed to characterize the spectra of an exo-Earth in the habitable zone.  An exo-Earth at 1 AU from the Sun has a contrast of 1.2$\times10^{-10}$ when the planet is at $90\degr$ phase angle (``half-moon''). We can select candidate stars by assuming a 1 $M_{\oplus}$ planet in the mid-habitable zone, 1 AU$\times\sqrt{L}$ (where L is stellar luminosity, in solar units), satisfying the following criteria: 1) star-planet contrast exceeds 4$\times10^{-11}$ at $90\degr$ phase, 2) star is brighter than $7^{th}$  magnitude, and 3) star is closer than 30 pc away. We have also eliminated binary stars from the candidate list. Figure ~\ref{fig:Figure3} shows the cumulative distribution of the number of target stars versus the maximum star-planet separation for a planet at mid-HZ. For example, there are 17 stars with maximum star-planet separation larger than 100 mas. There are a total of 296 stars that fit the above three criteria. The curve follows the power law $N\propto \theta^{-3}$ as expected (where N is the number of stars and $\theta$ is the angular separation between the star and planet), once we get past alpha Centauri A and B (two stars very close to the Sun where a terrestrial planet would be markedly easier to image). The list stops at 296 because of the $7^{th}$ magnitude and  4$\times10^{-11}$contrast cutoffs. In order to be able to search 60 stars for an exo-Earth in the HZ an imaging mission with minimum contrast 4$\times10^{-11}$ must have an IWA  of 65 mas or smaller, from Figure ~\ref{fig:Figure3}.

\subsection{The impact of IWA and contrast limit on orbit determination}

\begin{figure}
\includegraphics{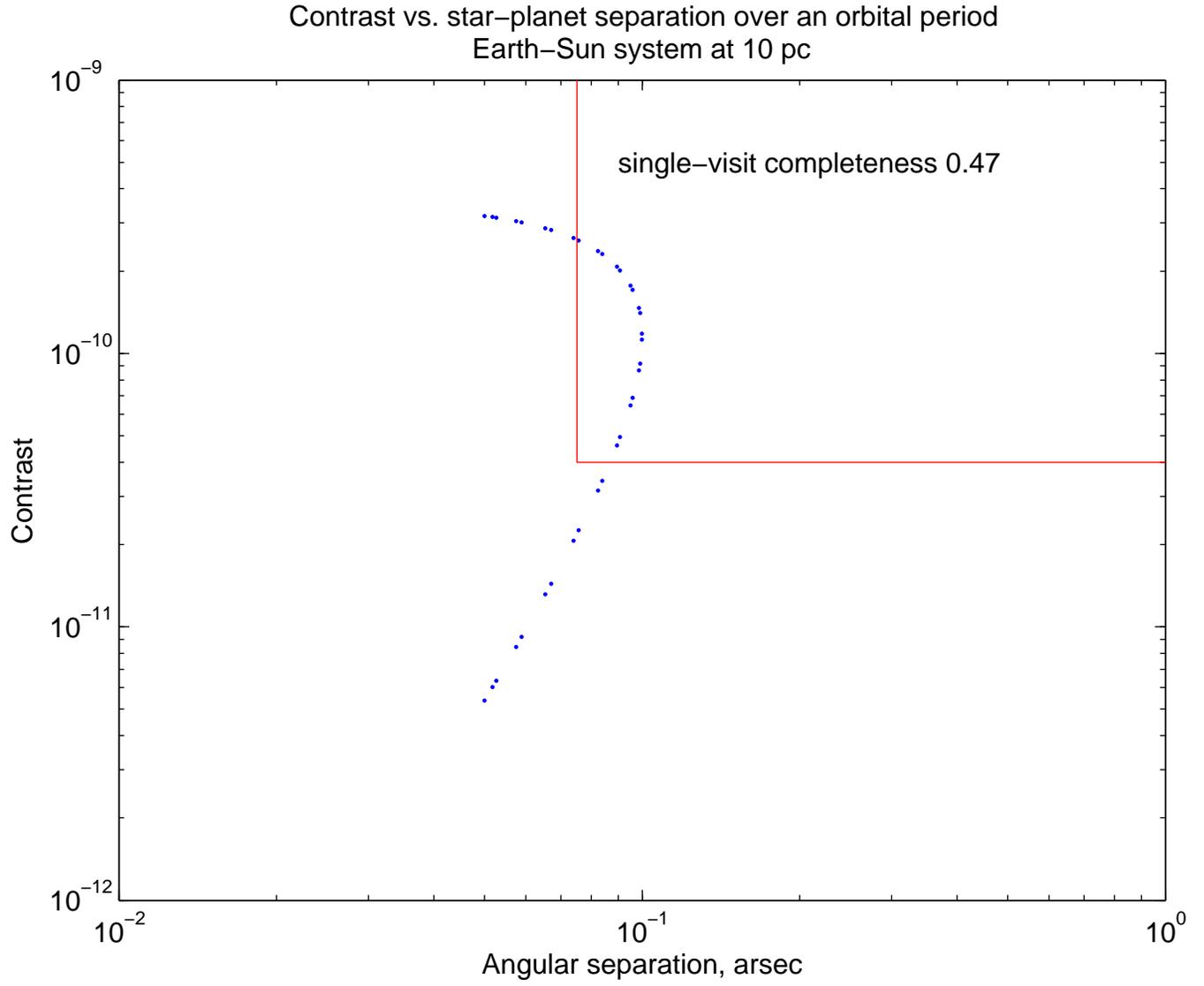}
\caption{Contrast and separation over an orbit, for an Earth-Sun system with $60\degr$ inclination at 10 pc. The left boundary of the red box is the inner working angle, and the lower boundary is the minimum detectable constrast. Only points inside the red box are observable. \label{fig:Figure4}}
\end{figure}

The minimum detectable contrast and the IWA limit the observable portion of the planetary orbit. Figure ~\ref{fig:Figure4} shows the contrast and star-planet separation for an Earth-Sun system at 10 pc as the planet orbits the star. The orbit is circular, with radius of 1 AU and $60\degr$ inclination (where $90\degr$ is edge-on). If the IWA is 75\% of the maximum star-planet separation and the minimum detectable contrast is 4$\times10^{-11}$ (3 times smaller than the contrast of Earth viewed at $90\degr$ phase angle), the planet would be observable over 47\% of its orbit. This number is called the single-visit completeness \cite{bro05}. Only the observations inside the red box are outside the IWA and exceed the contrast limit. This plot was generated assuming the planet is a Lambertian scatterer with an albedo of 0.3 and no seasonal dependent albedo.
For the same planet orbit and minimum detectable contrast, Figure ~\ref{fig:Figure5} plots the average number of images needed to see the planet 4 times (the minimum needed to confirm a planet's orbit with imaging alone) as a function of the ratio of IWA to maximum star-planet separation. In practice, if the IWA is only slightly smaller than the maximum star-planet separation, measurement of its orbit is impossible.

\subsection{Does astrometry help a direct imaging mission?}

This question has a couple of very different interpretations.  In one paper this question was interpreted as ``Can an astrometric orbit predict the location of a planet $\sim5$ years in the future with sufficient precision to guide a direct imaging mission as to when to look?''  We believe that a more relevant interpretation of the question is ``Does having an astrometric orbit or a null detection significantly help an imaging mission 1) identify which dot is which planet and 2) determine the direct imaging orbit, or 3) determine which targets not to look at?''

If a star has multiple planets, a prior astrometric mission will provide orbits for planets of Earth mass and larger that orbit  in the habitable zone and beyond. If two planets are detected in an image, a follow-up image in which either or both planets are seen will allow us to identify which planet is which, using knowledge of the orbits from astrometry. The same is true if the initial visit detects one planet and the follow-up image detects two.

For coronagraphic missions, the most important result from astrometry is the detection and non-detection of planets around the  target stars. If astrometry has detected a planet, we know that the probability that a planet exists is 99\%. Conversely, if an astrometric search of the HZ results in a null detection, it is very unlikely that there is an exo-Earth in the HZ.  For all of the top 60 imaging targets, SIM Lite (a space-based astrometry mission) could detect an exo-Earth at the inner HZ at a SNR of 5.8. Our Monte Carlo simulations show that if at a given star, there is no periodic signal above SNR 4.2 consistent with the HZ, then we can state with $\sim90\%$ confidence that there is no planet within the HZ that is more massive than Earth. If target stars that have a terrestrial planet in the habitable zone are identified prior to the imaging mission,  the imaging mission could concentrate on those stars. Knowledge of which target stars are unlikely to have a terrestrial planet in the  habitable zone saves a large fraction of the  mission time that would otherwise be wasted searching those stars for  planets.  For an occulter mission capable of spectral characterization, the proof-of-existence, masses, and orbits provided by an astrometric mission would free it to plan and optimize its resources for the function for which it is uniquely equipped, the characterization of planetary atmospheres.

The THEIA external occulter direct detection mission would visit  $\sim100$ targets to look for  exo-Earths, without measuring orbits or masses of the planets that it found. A  recent analysis considered whether a prior astrometric `classifier' mission that could determine whether or not there was a terrestrial planet in the habitable zone present at each target could help THEIA \cite{sa09a}. It was found that if $\eta_{\oplus} = 20\%$, THEIA alone would discover 7 of the 20 exo-Earths, but would find 13 if aided by knowledge from an astrometric `classifier' mission. The number of spectral characterizations between 250 and 1000 nm by THEIA alone was 4, but increased to 7 with the aid of the astrometric `classifier' mission. Evidently,  a prior astrometric `classifier' mission can significantly increase the science yield of the THEIA mission, when $\eta_{\oplus}$ is low. The analysis considered neither  the impact of the requirement to find orbits, nor the effect of confusion due to multiple planets. When these issues are considered, the advantage of a prior astrometry mission is even greater. In section 4, we analyze a small external occulter imaging mission to discover and measure orbits of exo-Earths. We include the effect of confusion due to multiple planets. We show that prior  astrometric detections of terrestrial planets in the HZ (at the 6 sigma level) would have a profound impact on the science return of this mission, if  $\eta_{\oplus}$ is between 10\% and 30\%.

Another argument that is often put forward is ``If exo-Earths are common,  e.g. every star has a 1 to 3 Earth-mass planet in the habitable zone,  we don't need astrometry to tell us which stars to avoid''.  But if we fly a small direct imaging mission, such as a 2 m-class coronagraph  that could only search the nearest 15 stars for an exo-Earth, then if $\eta_{\oplus}$ is 10\% and we have  no prior information from astrometry, there is a significant probability that zero exo-Earths are found. 

\section{Example: analysis of an external occulter imaging Mission}

In this section we compare the yield of an example imaging mission with and without prior astrometric information from SIM Lite.  For the imaging mission, we adopt the Occulting Ozone Observatory (O3) \cite{kas10}. External occulters are space-based coronagraph systems that use a specially shaped starshade occulter separated by a great distance from the telescope to suppress the starlight. They cannot look at stars closer than $45\degr$ from the Sun, or farther than $85\degr$ away from the Sun. The latter exclusion is because the Sun cannot illuminate the front of the starshade. The O3  mission concentrates on 30 target stars that are within 20 pc, brighter than $V = 5.5$ mag and have high single-visit completeness.  Before discussing the simulations and results,  we describe our approach to modeling planetary systems.

\subsection{Modeling realistic planet systems}
Any study of a planet-finding mission should evaluate the likely impact of a realistic distribution of planets on the science yield.  For example radial velocity (RV) studies have shown that multiple-planet systems are relatively common \cite{cum08}. Microlensing studies indicate that  super-Earth and Neptune mass planets (5 to 15 $M_{\oplus}$) are ~7 times more common than Jovian planets within a few AU of low-mass stars (~0.5 $M_{\bigodot}$) in the Galactic disk \cite{gou10}.
This is consistent with core accretion theory, which predicts that the mass probability distribution function rises toward lower masses \cite{mor09, ida08}. Radial velocity studies (sensitive to orbits faster than ~50 days) are uncovering a population of super-Earth and Neptune mass planets \cite{may09b, how09}. Core accretion models predict a peak in the planet population near 10 - 30 $M_{\oplus}$ composed of planets that did not manage to reach the stage of runaway gas accretion \cite{mor09}. For an astrometric mission such as SIM Lite, it's not significantly harder to find an exo-Earth in a multiple planet system than it would be if the exo-Earth  were alone \cite{tra10}. Current studies of direct imaging missions do not address the  problem of confusion due to multiple planets \cite{bro05, sa09a, sa09b, sav10}. A direct image of a planet measures the planet's projected position with respect to the parent star. A super-Earth, or a larger planet such as a Neptune-mass planet that is outside the habitable zone (HZ) could have a projected position that appears inside the HZ, and its flux contrast could be consistent with that of an exo-Earth that is closer in. For an example, see Figure 1 of \cite{cat10}. Any planet that appears at first detection to be a potential exo-Earth could therefore be a false alarm.

Radial velocity surveys have discovered over 400 planets to date \cite{exo10}. They reveal the mass and semimajor axis distribution and occurrence rate of gas giant planets.  Roughly 11\% of FGK stars have planets with masses between 0.3 and 10 Jupiters, with periods between 2 and 2000 days, and their distributions in mass and semimajor axis are consistent with being uniform in log space. Systems with multiple gas giant planets are relatively common -- 41 out of 351 (or 12\%) of the planetary systems discovered by RV have more than one planet.  RV surveys have begun to probe a new population of ice giant planets, with masses between a few and 100 $M_{\oplus}$ \cite{may09b}. Microlensing surveys, with more limited statistics, indicate that ice giants within a few AU (cold neptunes) are common. They are estimated to occur at a rate of 0.36 per $dex^{2}$ in the phase space of mass and semimajor axis (a dex is a decade interval in the logarithmic domain). This gives about 11\% probability of finding a Neptune-mass planet in the HZ.  Exo-Earths, the objects of our search, are contained in a third population, terrestrial planets, with masses between 1 and 10 $M_{\oplus}$. A few planets in this mass range have recently been found, by RV and via their transits by Corot and Kepler. The mass and semimajor axis distribution and occurrence rates of gas giants are known; for ice giants we have estimates based on microlensing and RV data, but for terrestrial planets we have to make an educated guess. Information from Kepler will eventually help determine the occurrence rate and mass and semimajor axis distributions of terrestrial exoplanets. In the absence of this information, it would seem reasonable to assume that the distribution of their masses and semimajor axes are also uniform in log space. For consistency with previous studies, we adopt values of 10\% to 30\% for $\eta_{\oplus}$, the fraction of stars with planets with masses between 1 and 10 $M_{\oplus}$ that are within the HZ.

To model a planetary system, we draw planets from two populations: ice giants (10 to 100 $M_{\oplus}$), and terrestrial planets (1 to 10 $M_{\oplus}$), according to their occurrence rates and mass and semimajor axis distributions. We assume circular orbits with radius ranging between 0.1 and 10 AU. Orbital inclinations (assumed the same for all planets in a system) are drawn from a uniform distribution in cosine(inclination), and orbital phase for each planet is randomly drawn on the interval $[0,360\degr]$.

We draw the ice giant planet from a population consistent with the results of microlensing \cite{gou10}, and RV \cite{may09a}surveys. The planet density is is assumed to be uniform in the phase space of log(mass) and log(semimajor axis), with the value $\frac{dN}{dlogMdloga}=0.36$ planets per $dex^{2}$. 

Integration of the planet density inside the region bounded by 10 and 100 $M_{\oplus}$ in mass, and 0.1 and 10 AU in semimajor axis yields 0.72 planets per star, which means there is a 72\% chance that a star has a planet with mass and semimajor axis in this range. We draw a random number between  0 and 1,  and if it's less than  0.72  the star gets an ice giant planet.  To assign the planet's mass and semimajor axis, we draw a random number from a uniform distribution on the interval [1, 2] for log(mass) and a random number from a uniform distribution on the interval [-1, 3] for log(semimajor axis).

For the exo-Earth population, we again assume that  the distribution is uniform in log(mass) and log(semimajor axis). The region of phase space for terrestrial planets is bounded by 1 and 10 $M_{\oplus}$ in mass, and by 0.1 to 10 AU in semimajor axis. The integrated planet density over the habitable zone (between 0.8 AU and 1.6 AU) is set equal to the chosen value of $\eta_{\oplus}$. With this normalization, the value of the (constant) planet density per $dex^{2}$ is determined.  Integrating the planet density over the region of interest in phase space gives planet occurrence rates of 0.66, 1.33 and 1.99 per star for $\eta_{\oplus}$ = 0.1, 0.2 and 0.3, respectively.

For eta-earth = 0.1, we draw a random number from a uniform distribution on [0,1]; if the number is less than 0.66 we draw one planet from this population.

For $\eta_{\oplus}$ = 0.2 and 0.3 we draw one planet from this population, and use the following procedure to decide whether to draw a second planet:

Draw a random number from a uniform distribution on [0,1]
\begin{itemize}
  \item If $\eta_{\oplus}$ = 0.2, draw a second planet if the random number is less than 0.33.
  \item If $\eta_{\oplus}$ = 0.3, draw a second planet if the random number is less than 0.99.
\end{itemize}

We considered also drawing a planet from the known distribution of Jupiter-class planets \cite{cum08}, but it turns out that only ~2.3\% of stars have planets of mass between 100 $M_{\oplus}$  and 10 $M_{Jup}$ in the HZ. Since the Jupiters didn't much affect the results of the simulations, we decided to omit them.

\subsection{Occulting Ozone Observatory (O3) survey, with no prior knowledge from astrometry}

The Occulting Ozone Observatory (O3) is a small externally occulted telescope with a 1.1 m mirror that is designed to survey 30 stars for exo-Earths. It collects light in visible and UV wavebands, which allows it to search for the presence of atmospheric ozone.   Following the description of the O3 mission \cite{kas10}, we simulated a survey of the top 30 occulter targets (D. Lisman, private communication) ranked by completeness, drawing planets for each one randomly from the three populations, as described in section 4.1. We note here that 11 of their top 30 targets are binaries, and the stellar separations may dynamically exclude planets in the habitable zones. The top 30 occulter targets have a median  single-visit completeness of 58\%. An initial survey visits each of these stars once. In the first visit, either a planet, several planets, or no planet is detected. If a planet is detected with a projected position appearing inside the outer HZ for the star, and its flux contrast is below the limit for a maximum mass (i.e. maximum radius) terrestrial planet at the inner HZ,  it is considered a potential exo-Earth.

The mass-radius relationship for solid planets is subject to strong degeneracies and the mass at a given radius is quite sensitive to the assumed structure and composition of the planet \cite{sea07}. For the simulations in this study, we have adopted a simple exo-Earth model with mass proportional to radius cubed, normalized to the mass of the Earth. To estimate the upper bound to the  flux contrast of a terrestrial planet, the radius of a 10 $M_{\oplus}$  planet is therefore taken to be 2.15 $R_{\oplus}$. If the planet's flux contrast is too high then it's not in the terrestrial mass range, and if its projected position is outside the HZ boundary then it isn't in the HZ.  A followup survey returns to each of the best 6 candidate stars for 5 additional observations spaced apart in time by $P_{HZ}/3$, where $P_{HZ}$ is the period of a planet orbiting at 1 AU radius, scaled by $\sqrt{L}$. The highest followup priority goes to stars whose initial survey observation revealed the image of a potential exo-Earth.  If there are fewer than 6 stars with potential exo-Earths, the remainder of the 6 followup stars are chosen from the highest-completeness stars not already on the followup list. If no potential exo-Earth is detected on the first visit to a  target, it will most likely not be visited again, even though it may still harbor a potential exo-Earth.  The total of 6 observations per candidate star ensures that if there is a terrestrial planet in the HZ, it will be detected on average 4 times, which is enough to confirm the planet and solve for the orbit. There are a total of $60 = 30 + 6\times5$ observations in the survey.

We considered two scenarios:  (1) Allowing only planets that are of terrestrial mass and orbiting within the HZ (as assumed by most direct imaging studies to date) and (2) Adopting more realistic planet mass and semimajor axis distributions, as discussed in section 4.1. For each scenario, we ran 1000 survey realizations for each of the cases $\eta_{\oplus} = 10$\%, 20\% and 30\%.
The results of our simulations are presented in Table ~\ref{tbl-1}. Note that for the simulations presented in this table, we used the O3 target list.

If we restrict the planets to be of terrestrial mass and inside the HZ, our occulter simulations find a mean of 1.7, 3.1, and 4.2 exo-Earths, for the cases of $\eta_{\oplus} = 0.1$, 0.2 and 0.3 respectively. This is roughly consistent with the result of the O3 group, in which their simulated occulter survey found 5 exo-Earths for $\eta_{\oplus} = 0.3$ \cite{kas10}.  With the more realistic distribution of planet masses and semimajor axes simulated in scenario (2), the occulter survey was significantly less productive, yielding only 1.6, 2.7, and 3.3 exo-Earths, for $\eta_{\oplus}= 0.1$, 0.2 and 0.3 respectively.

If $\eta_{\oplus}= 10\%$ the imaging mission could only expect to find 1.6 planets (on average), and there was  a probability of 17\% that it would find no planets. Why does scenario (2) have a lower yield of exo-Earths than scenario (1)?  Allowing planets with the full range of masses and orbit radii introduces the possibility that the initial detection may be a false alarm. There are roughly $\log(100)/\log(2) = 7$ times more terrestrial planets in any mass range orbiting between 0.1 and 10 AU than in the HZ (0.8 to 1.6 AU). There are also Neptune-mass planets: 72\% of the target stars have a Neptune-mass planet between 0.1 and 10 AU, and 11\% have a Neptune-mass planet in the HZ. In this more realistic environment, confusion takes its toll; some planets detected in the initial survey will be Neptune-mass planets whose flux contrast and projected position are consistent with a rocky planet in the HZ, and some will be terrestrial planets that are not in the HZ, but whose projected position is within the HZ.  These false alarms will not pan out in the followup survey.

In a related paper, we presented a similar study of a nominal JWST + external occulter mission to survey 26 targets for exoplanets \cite{cat10}. The results of the orbit measurement survey for the JWST mission are somewhat  poorer than for O3. The main reasons are that JWST operates in the near infrared, where Earth's geometric albedo is 30\% smaller than in the optical, and the JWST mission uses a more restrictive target star list.

If a star has a Neptune and an exo-Earth within a few AU, knowledge of the Neptune's orbit from an RV survey together with N-body simulations can indicate whether there are niches of stability that could allow an exo-Earth to exist within the habitable zone \cite{mal07}.  But previous knowledge of the Neptune's orbit doesn't help when it comes to sorting out whether an image of a planet is a Neptune or an exo-Earth.  If we know a star has a Neptune-mass planet within a few AU, the star could have an Earth at 1 AU, so it is still a viable target. An orbit from RV does not predict where the Neptune-mass planet will be in an image, because it does not give complete information; the azimuthal orientation of the semimajor axis with respect to  the star  and the orbital inclination remain unknown. In an image, the Neptune-mass planet could have a flux and apparent position  consistent with an exo-Earth in the habitable zone, thereby causing a false alarm.

For imaging missions, realistic assumptions about planet mass and semimajor axis distributions introduce confusion between exo-Earths and non-exo-Earths that has not been considered in previous studies \cite{bro05, sa09a, sa09b, sav10}, where only the presence or absence of exo-Earths is envisioned. Our results (see Table 1) show  that  this confusion significantly reduces the yield of an imaging mission.  As $\eta_{\oplus}$  grows larger, the relative confusion from the terrestrial-mass planets stays the same, since it scales with  $\eta_{\oplus}$, but the confusion from Neptune-mass planets is diluted, since their number stays constant.

Figures ~\ref{fig:Figure6} \&~\ref{fig:Figure7} show an example of a false alarm caused by a Neptune-mass planet outside of the HZ. This system has 2 terrestrial mass planets  and a Neptune-mass planet.  In these figures, red denotes the Neptune-mass planet, cyan and black denote the terrestrial mass planet; a filled circle is a detection, an empty circle is a miss.

Figure ~\ref{fig:Figure6} shows the positions of the planets at the times of observation, and Figure ~\ref{fig:Figure7} shows the fluxes. Since   the fluxes and apparent positions of the Neptune-mass planet are consistent with a terrestrial planet in the habitable zone,  it would be a false alarm. The orbits of the two terrestrial planets are unobservable because they are inside the IWA.

\subsection{Occulter Ozone Observatory (O3) survey with prior knowledge from astrometry}

We next consider the occulter mission aided by prior knowledge from a space-based astrometry mission such as SIM Lite. SIM Lite has very nearly 100\% completeness and reliability for rocky planets in the HZ. SIM Lite can survey the best 60 occulter targets for Earth mass planets in the habitable zone. For the top 60 stars in the occulter list, SIM Lite will find the planets and measure their orbits and masses. This prior knowledge significantly improves the science yield of the occulter mission.

The occulter would need only to confirm the detection of a planet in order to refine the orbit and mass estimate. With targets averaging 47\% completeness, it takes on average two tries to find the planet the first time,  which nails down the orbital phase. A third observation confirms and further refines the orbit. Given 60 observations, the occulter could follow up a maximum of 20 SIM Lite discoveries.  For $\eta_{\oplus} =$ 0.1, 0.2 and 0.3, there would be 6, 12, and 18 stars with exo-Earths among the top  60 occulter targets. With prior knowledge from SIM Lite, 95\%of these would be known, and could be characterized by the occulter mission (with well under 60 observations for the first two cases), and the remaining mission time could be used to follow up SIM Lite discoveries at lower-priority occulter targets.  For the example of the O3 occulter mission, prior knowledge from an astrometric mission such as SIM Lite would increase by a factor of $\sim$4 to $\sim$5 the number of exo-Earths that could be discovered, if $\eta_{\oplus}$  were between 0.1 and 0.3 (Refer to Table 1).

\section{Internal coronagraphs, and an example}

\begin{figure}
\includegraphics{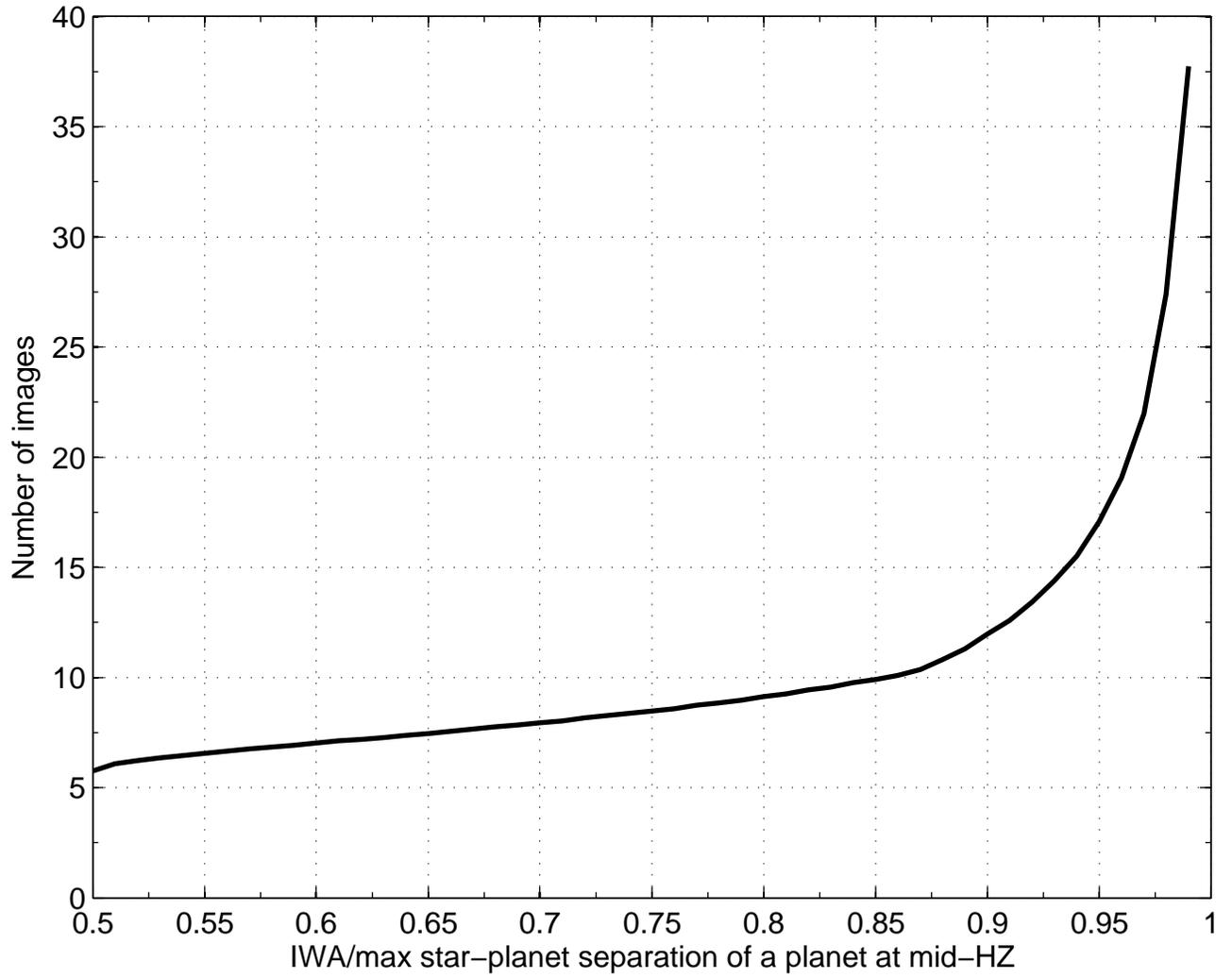}
\caption{The number of images needed to detect a planet 4 times so as to measure and confirm the orbit.  When the maximum star-planet separation is close to the IWA, the completeness is low, so more images are required.\label{fig:Figure5}}
\end{figure}

\begin{figure}
\includegraphics{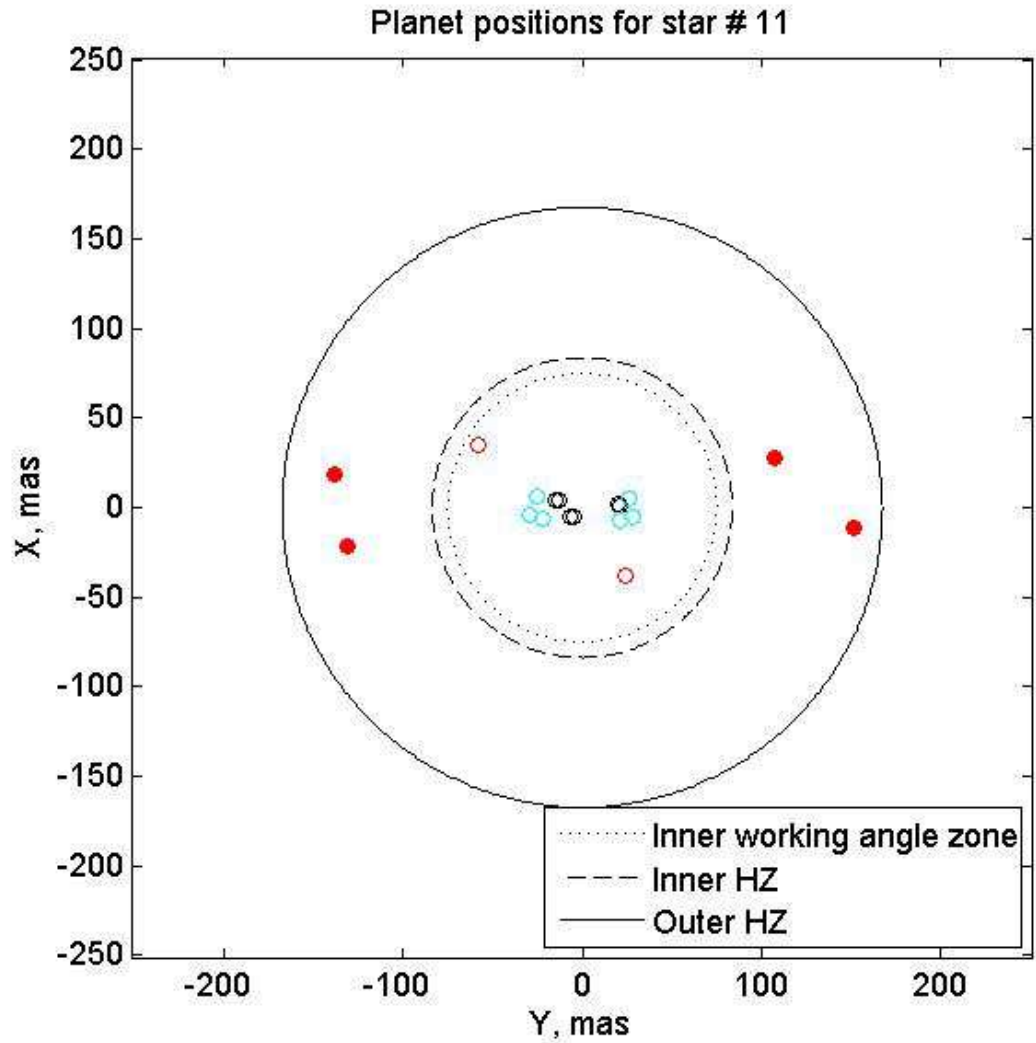}
\caption{Positions of planets at the epochs of observation. Red, black, and cyan circles are the positions of a Neptune-mass planet, a terrestrial mass planet, and a second terrestrial mass planet. Filled circles are detections, unfilled circles are non-detections.\label{fig:Figure6}}
\end{figure}

\begin{figure}
\includegraphics{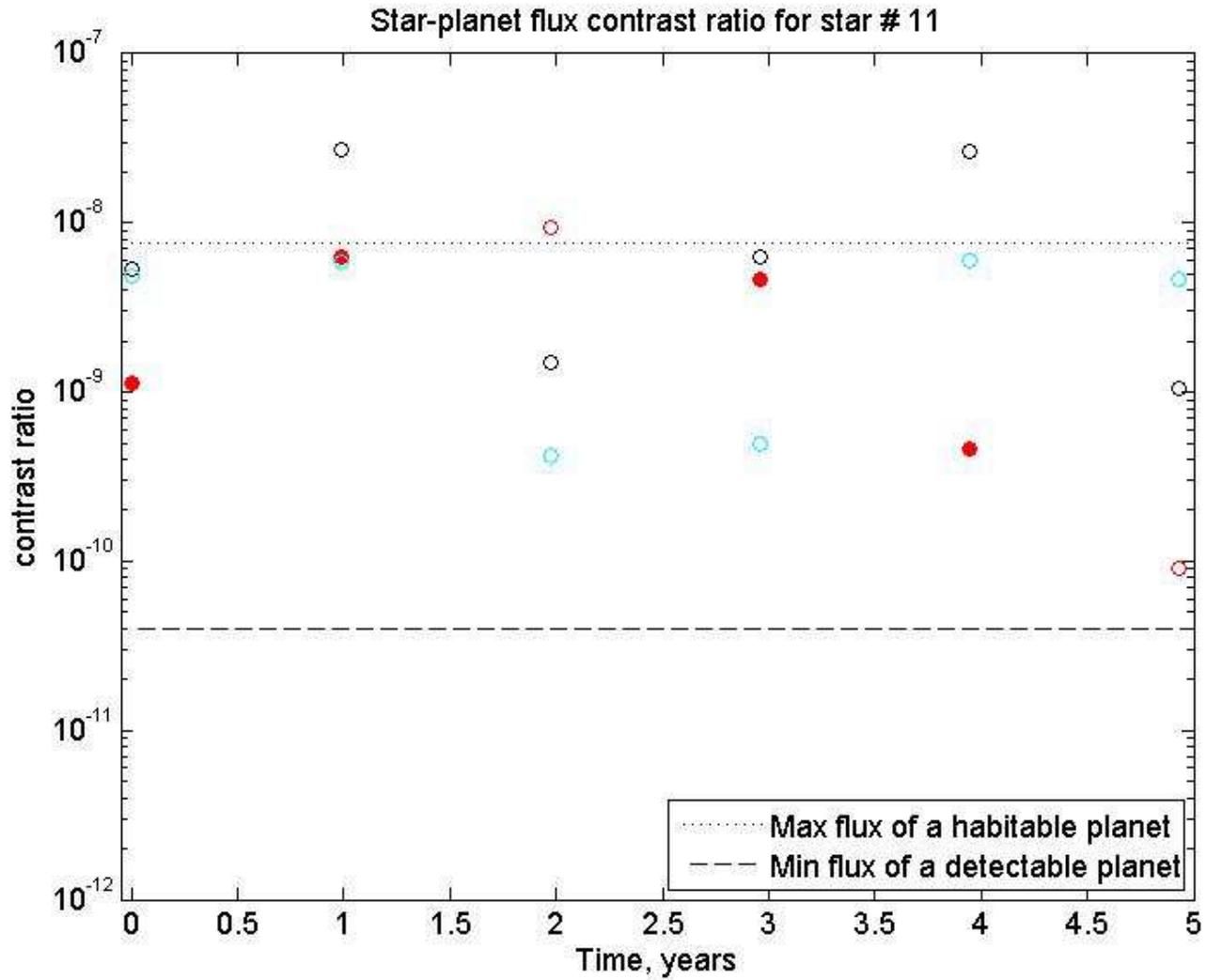}
\caption{Planet-star flux contrast ratios for the planets at the epochs of observation. Red, black, and cyan circles are the flux contrast ratios of a Neptune-mass planet, a terrestrial mass planet, and a second terrestrial mass planet. Filled circles are detections, unfilled circles are non-detections.\label{fig:Figure7}}
\end{figure}

Internal coronagraphs do not have nearly as serious a limitation on the number of visits as do external coronagraphs, because slewing between targets does not occupy such a large fraction of mission time. The number of visits is limited only by the total mission lifetime and the integration time needed to detect an exo-Earth. Internal coronagraphs can access more of the sky at any epoch, since they do not have the added solar exclusion constraint imposed by the starshade. However, internal coronagraph missions often include stars in the list of potential targets where a planet in the habitable zone would result in a maximum star-planet separation barely larger than the IWA. Measuring an orbit requires 4 detections, and if one wants to get 4 detections in 6 tries, the first-visit completeness has to be $\sim60\%$, requiring the maximum star-planet separation to be larger than $\sim2\times$IWA (See Figure ~\ref{fig:Figure5}). This factor of 2 in the effective inner working angle becomes a factor of 8 in the volume of space and number of potential targets excluded.   A program that aims to take 4 images of every planet should adjust the number of visits according to the imaging completeness.

If the desired number of potential targets is 60 to 100, we now have to look at targets 15 or even 20 pc away, not just stars at a distance of 5 to 6 pc. Astrometric detection of an exo-Earth requires an integration time that goes as the square of the distance to the target.  Direct detection of an exo-Earth is background-limited by the local and exo-zodi. It therefore requires an integration time that goes as the 4th power of the distance. A moderate-sized internal coronagraph that can detect an exo-Earth around Alpha Centauri (1.3 pc) in 1 minute of integration will require a week for that same planet at 13 pc and 39 days when that planet is at 20 pc.

The original TPF-C mission with an 8 m telescope working at 4$\lambda/D$ had an IWA $\sim60$ mas \cite{lev09}. There were over 130 nearby stars where $R_{max}$, the maximum star-planet separation for a planet in the middle of the habitable zone, was greater than 60 mas. If such a coronagraph were to visit each star just once and observe long enough to get a SNR=12 image of a 1$M_{\oplus}$ planet, the total integration time was much less than a typical 5 year mission. But a survey of 100 nearby stars that would also get the orbits of the potential exo-Earths would require much longer integration times.

The survey of 30 stars by the O3 mission assumed that sufficient time was spent to detect a planet with a contrast of 4$\times10^{-11}$; an Earth at 1 AU and $90\degr$ phase angle has a contrast of
1.2$\times10^{-10}$. If we wish the detection to have a false alarm probability of 1\%, the speckles have to be subtracted to $\sim1/12$ of 4$\times10^{-11}$ (see Appendix). In other words, the 1 sigma noise in the image has to be 3.3$\times10^{-12}$ of the star. A survey of stars to 4$\times10^{-11}$ with 1\% FAP requires 16 times longer integration per star than a SNR=8 image of an exo-Earth.

\begin{figure}
\includegraphics{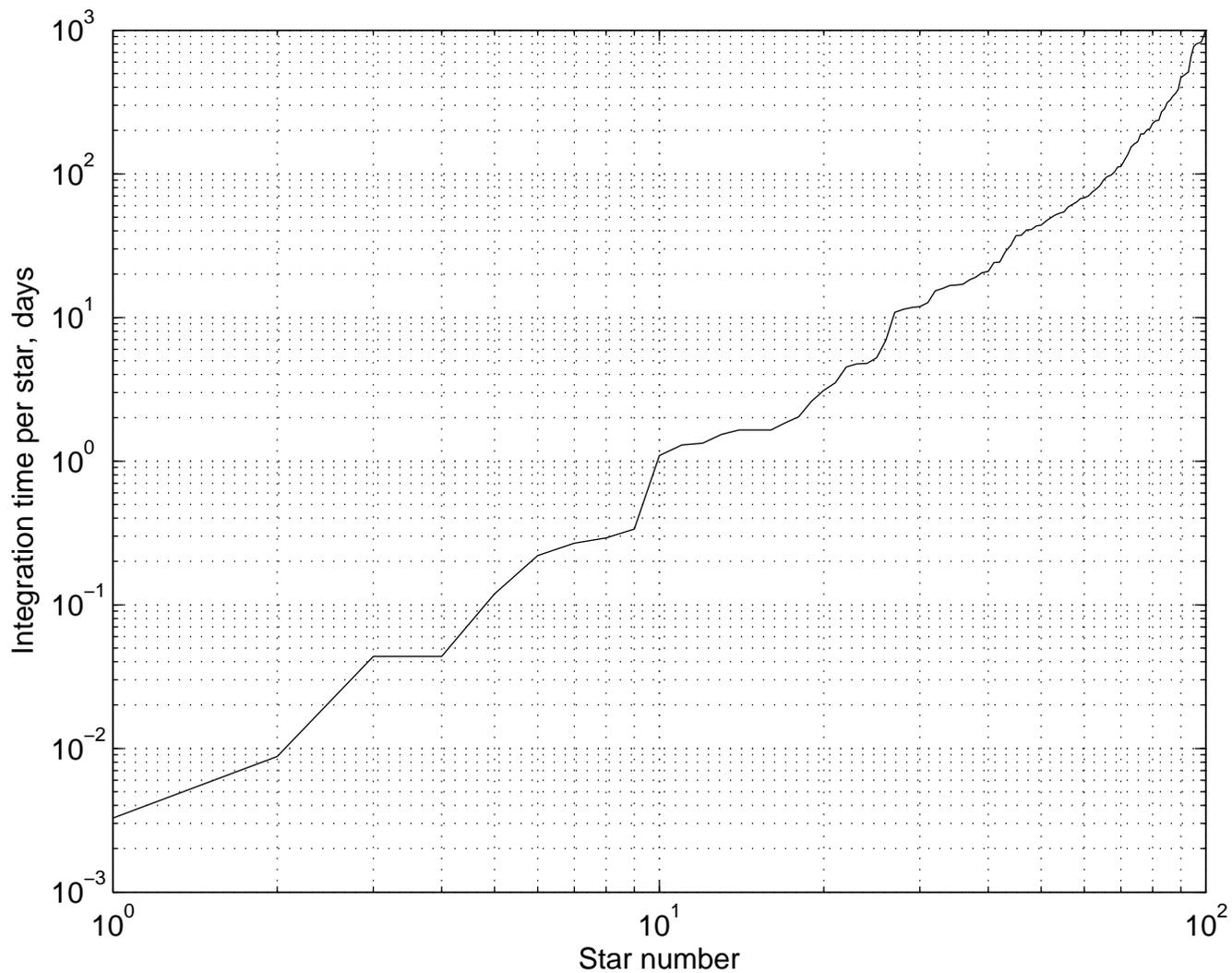}
\caption{Integration time needed for each of the best 100 imaging targets, to  measure and confirm an orbit (4 detections) for an exo-Earth at the mid-HZ. The total integration time needed for the best 60 stars is 3.3 years, and for the best 100 stars is 39.2 years. See table ~\ref{tbl-2} for the assumptions in the integration time calculation.\label{fig:Figure8}}
\end{figure}

The number of visits needed to image an exo-Earth 4 times depends on the ratio $IWA/R_{max}$ as shown in Figure ~\ref{fig:Figure5}. As an example, we calculated the on-target integration time for an internal coronagraph with a D = 3.5 m diameter unobscured telescope with a 2$\lambda/D$ IWA of 70 mas, and a total efficiency (QE and optics losses), of 25\%. The integration time for a detection is derived from the SNR as a function of integration time:
\begin{equation}
SNR=\frac{Pt}{\sqrt{(P+B_{Z}+B_{S}+B_{D})t+RN} }
\end{equation}
In this equation, P is planet flux, t is integration time, $B_{Z}$ is zodi background, $B_{S}$ is speckle background (see Appendix),  $B_{D}$ is dark current background, and RN is read noise.
The assumptions in the integration time calculation are in Table ~\ref{tbl-2}.
The result of the calculation is shown in Figure ~\ref{fig:Figure8}.

Figure ~\ref{fig:Figure8} shows the integration time needed for each of the ``best'' 100 nearby stars, to detect an Earth-like planet 4 times, with a sensitivity down to $1/3$ of the flux of a 1 Earth mass planet at $90\degr$ phase angle. For the nearest stars where $IWA/R_{max}$ $\sim0.5$, only 5 or 6 attempted images would be needed to see the planet 4 times. For the more distant stars, up to $15\sim20$ attempted images may be needed. The ability to detect planets down to a contrast of 4$\times10^{-11}$ means that one could find Earths around late F stars, and even for G stars would significantly increase the first-visit completeness by making the planet detectable even when the bright side of the planet was slightly facing away from the coronagraph.

The cumulative integration time needed to detect an exo-Earth 4 times if it existed, is, for the best 60 stars, $3.3$ years. The integration time needed to get orbits for the 100 best stars is  39.2 years. If $\eta_{\oplus}$ is 10\%, and the existence of the $6\sim10$ exo-Earths were already known and identified by astrometry, the coronagraph would save a factor of ten in search time.  A further factor of 2 is realized in imaging time,  because only two images of the planet are needed if the astrometric orbit is known, versus four images without prior astrometry.  The coronagraph could confirm the orbits in 1/20 of the above listed times. The 10 Earths around the best 100 stars would take $\sim2$ years.  A prior astrometric mission  would mean roughly the difference between a 4 and  8 meter coronagraph.

\section{Summary and conclusions}

In this paper we have simulated the planet-finding capability of the Occulting Ozone Observatory (O3), with and without a prior astrometric mission.  O3 is an imaging mission which uses a 1.1 m telescope plus an external starshade occulter to measure orbits of exo-Earths. In our analysis, we used the same target list as in \cite{kas10}. We simulated realistic planet distributions, which allow multiple-planet systems. This study is the first that we are aware of to account for the confusion that can arise among exo-Earths, super-Earths, and Neptune-class planets for an imaging mission. In the near future, RV and microlensing data will further constrain the Neptune and super-Earth populations, and Kepler data will reveal the statistics of exo-Earths, allowing us to improve the realism of our planet population models. In this work, we have assumed circular planet orbits, and we have assumed that 4 images are sufficient to fit an orbit. In future work, we will include eccentric orbits, and we will model the detection process by actually fitting the planet orbits.

A planet's orbit can be measured by space astrometry or by multiple direct images, or a combination of the two.  Prior astrometry helps in two important ways. If an exo-Earth has been discovered and its orbit measured astrometrically, the orbital phase extrapolated to 5 years in the future would have a sizable error bar.  However, as we have shown in this paper, given an orbital solution from a prior astrometric mission, the orbital phase can be pinned down very accurately by a single follow-up image. Without this knowledge, a coronagraph would need a much larger number of images to discover the planet and measure the orbit (see Figure ~\ref{fig:Figure5}). Secondly, knowing beforehand which stars do not have exo-Earths would save the many hours of coronagraph search  time at these stars that would be necessary to make such a determination. This is potentially a large fraction of mission time, which could  become available for spectroscopy of the exo-Earths, or for more orbit measurements. With information from prior astrometry, the coronagraph gains efficiency; it could survey more targets for exo-Earths in the same mission time.

The impact of astrometry is most pronounced if the follow-on direct detection mission is an external occulter. Small occulters might be limited to $\sim60$ pointings, large occulters to $\sim120$ pointings. A small occulter (e.g. the O3 mission) as a follow-on to an astrometric mission could detect  $\sim6$ exo-Earths, if $\eta_{\oplus} = $10\%. By itself, this small external occulter would detect  just $\sim1.5$ exo-Earths, with a sizeable probability that zero are detected. Small internal coronagraphs that only have $\sim10$ potential targets would spend a large fraction of their mission visiting each of these targets 10 to 15 times each.
For internal coronagraphs, there is no propellant limitation and if one builds a large enough telescope, there will be sufficient integration time to survey 60 to 100 nearby stars. When the target list extends to 100 stars,  a survey of all 100 stars with 4 images of the planet at a contrast of 4$\times10^{-11}$  would take 40 years.  But knowledge of which stars have Earths and which do not could reduce the integration time by roughly a factor of 20. The factor of 20 corresponds to slightly more than a factor of 2 in the diameter of the coronagraphic telescope.

When it comes time to build a large internal coronagraph, knowing where the exo-Earths are will make it possible to design a mission just large enough to do what is scientifically compelling, rather than a mission that is much more expensive, to insure against bad luck.

\acknowledgments
The research described in this paper was carried out at the Jet Propulsion Laboratory, California Institute of Technology, under contract with the National Aeronautics and Space Administration. Government sponsorship acknowledged.  We thank Doug Lisman for providing the target list for the O3 mission, and Dmitri Savransky for information about the O3 mission. We thank John Davidson,  Steve Edberg, Doug Lisman, Stuart Shaklan and Steve Unwin for helpful discussions and comments. We are grateful to the anonymous referee for providing numerous useful suggestions and comments that  helped improve the paper.

\appendix

\section{Appendix: Detecting planet in a sea of speckles}

A high contrast coronagraph blocks out the light from a star so that the light from a planet orbiting the star can be seen. If the coronagraph had perfect optics, or the external occulter
was manufactured perfectly and aligned perfectly, 100\% of the starlight would be blocked, leaving only the planet light. But very small imperfections in the optics can cause ``speckles'' to appear in the image. Speckle intensity is proportional to the square of the wavefront error.  With an internal coronagraph, a sinusoidal wavefront error of rms amplitude 1 picometer (pm) would produce a speckle of intensity $10^{-10}$ of the star. By comparison, in an Earth-Sun system the planet at $90\degr$ phase (half-illuminated) would have a brightness of 1.2$\times10^{-10}$. Building large telescope optics to $10^{-6}$ wave tolerance is well beyond current technology. Coronagraphic concepts instead use deformable mirrors (DM's) to remove the wavefront errors in the telescope and instrument optics. A 1000 element DM $\sim32\times32$, could control spatial frequency up to $\pm16$ cycles/aperture, and remove speckles in a $32\times32$ Airy spot region centered on the star. If the DM is set with an accuracy of $\sim30$ pm rms, the average energy at one sinusoidal frequency in the pupil corresponding to one speckle location in the image plane would be 1 pm rms.

Control of the wavefront to 30 pm rms would therefore produce a sea of speckles with average flux of $\sim10^{-10}$.  If the average speckle's flux is $10^{-10}$, and the planet's flux is $10^{-10}$, then half the speckles would be brighter than the planet and half would be dimmer. If we set a detection threshold of $10^{-10}$, there would be 500 false alarms in a field of view of 1000 pixels, where each pixel is of angular size $\lambda/D$.  To reliably detect a planet with $10^{-10}$ contrast, the average speckle flux must be significantly lower than $10^{-10}$.  We adopt a metric used in indirect detection, called false alarm probability. To detect a planet in a sea of speckles, we want the false alarm probability (FAP) to be 1\% per image. If we assume the 30 pm wavefront errors are random, with a Gaussian distribution, it's straightforward to calculate the probability distribution of speckle brightnesses.
The intensity of the speckles obeys an exponential distribution. Suppose that the wavefront, after correction by one or more DM's, has an rms phase error of 30 pm and an rms amplitude error over the pupil of 0.03\%. Then the speckles have a mean intensity of $2\times10^{-10}$, which includes equal contributions from wavefront amplitude and phase errors, and the standard deviation of the speckle intensity is also 2$\times10^{-10}$. If we take the imaging field of view to have 1000 independent pixels (each $\lambda/D$ across) and we want the FAP to be 1\%, then the probability of any one speckle being larger than the threshold of detection has to be 1$\times10^{-5}$. This is achieved when the detection threshold is set to $\sim12$ times the mean speckle brightness. That is, for speckle statistics, a SNR of 12 is needed for detection of a planet with 1\% FAP. Detection of a planet with flux contrast of $10^{-10}$ would then require the average speckle brightness to be $\sim8\times10^{-12}$. On the other hand, if the image is dominated by photon or exo-zodi or detector noise, rather than speckle noise, a SNR of 5 is all that is needed for 1\% FAP detection. In general, the photon noise contribution must satisfy the SNR = 5 condition, and the speckle noise contribution must satisfy the SNR = 12 condition.

Because a contrast of $8\times10^{-12}$ is much more difficult to achieve than a contrast of $10^{-10}$, almost all coronagraphic approaches use some form of post-detection speckle removal. If the speckles are very stable, one might rotate the telescope around the direction to the star. The speckles will rotate, but the planet will not. This approach to speckle subtraction replaces ``accuracy'' of the wavefront at the single digit picometer level with ``stability'' of the wavefront at the single digit picometer level.  Detection of a $10^{-10}$ planet with 1\% FAP requires speckles (after PSF subtraction) to be $\sim8\times10^{-12}$ on average, implying that an initial wavefront error of 30 pm rms be stable to $\sim2.5$ pm. However, in most coronagraphic mission simulations, the assumption is that the coronagraph can detect a planet that is $4\times10^{-11}$, not $1.2\times10^{-10}$ as bright as the star. This requires the average speckle brightness to be below $3.3\times10^{-12}$. This additional factor of $\sim3$ contrast means that an AO coronagraph's initial wavefront error of 30 pm rms needs to be stable to slightly less than 1 pm rms. If the speckle subtraction is performed by rotating the telescope after an hour of integration, the required stability is roughly 1 pm/hr.


\clearpage
\begin{table}
\begin{center}
\caption{Mean number of exo-Earths detected and characterized by the O3 occulter mission\label{tbl-1}}
\vspace{0.07in}
\begin{tabular}{lccc}
\tableline\tableline
Mission & $\eta_{\oplus}=10\%$ & $\eta_{\oplus}=20\%$ & $\eta_{\oplus}=30\%$ \\

\tableline
Imaging survey, scenario 1:\\   terrestrial planets only in HZ & 1.7 & 3.1  & 4.2 \\
\tableline
Imaging survey, scenario 2:\\ realistic planet distribution & 1.6 & 2.7 & 3.3 \\
\tableline
Imaging survey with prior \\
knowledge from SIM Lite, scenario 2:\\
 realistic planet distribution & 6 & 11 & 17 \\

\tableline
\end{tabular}
\end{center}
\end{table}

\begin{table}
\begin{center}
\caption{Assumptions for imaging time calculations\label{tbl-2}}
\vspace{0.07in}
\begin{tabular}{ll}
\tableline
\tableline
Local (solar) zodi level   &	Average at $45\degr$  ecliptic latitude 22.7 mag arcsec$^{-2}$\\
\tableline
Exozodi level    &	2 solar zodi @ $45\degr$ exo-ecliptic latitude\\
\tableline
Detector dark current	&0.001 electrons sec$^{-1}$ pixel$^{-1}$\\
\tableline
Number of pixels in matched filter & 	23\\
\tableline
Pixel size                 &	$0.5\lambda/D$  \\
\tableline
Read noise                 &	2 electrons \\
\tableline
Total telescope/coronagraph throughput  & 	25\%\\
\tableline
IWA   	&  $2\lambda/D = 70$ mas \\
\tableline
Telescope Diameter	&3.5 m  \\
\tableline
SNR for a planet of 1/3  Earth\\flux contrast at quadrature   &12\\

\tableline
\end{tabular}
\end{center}
\end{table}

\end{document}